
%
%
%
\font\twelverm=cmr10 scaled 1200    \font\twelvei=cmmi10 scaled 1200
\font\twelvesy=cmsy10 scaled 1200   \font\twelveex=cmex10 scaled 1200
\font\twelvebf=cmbx10 scaled 1200   \font\twelvesl=cmsl10 scaled 1200
\font\twelvett=cmtt10 scaled 1200   \font\twelveit=cmti10 scaled 1200
\font\twelvesc=cmcsc10 scaled 1200  
\skewchar\twelvei='177   \skewchar\twelvesy='60


\def\twelvepoint{\normalbaselineskip=12.4pt plus 0.1pt minus 0.1pt
  \abovedisplayskip 12.4pt plus 3pt minus 9pt
  \belowdisplayskip 12.4pt plus 3pt minus 9pt
  \abovedisplayshortskip 0pt plus 3pt
  \belowdisplayshortskip 7.2pt plus 3pt minus 4pt
  \smallskipamount=3.6pt plus1.2pt minus1.2pt
  \medskipamount=7.2pt plus2.4pt minus2.4pt
  \bigskipamount=14.4pt plus4.8pt minus4.8pt
  \def\rm{\fam0\twelverm}          \def\it{\fam\itfam\twelveit}%
  \def\sl{\fam\slfam\twelvesl}     \def\bf{\fam\bffam\twelvebf}%
  \def\mit{\fam 1}                 \def\cal{\fam 2}%
  \def\sc{\twelvesc}               \def\tt{\twelvett}
  \def\sf{\twelvesf}
  \textfont0=\twelverm   \scriptfont0=\tenrm   \scriptscriptfont0=\sevenrm
  \textfont1=\twelvei    \scriptfont1=\teni    \scriptscriptfont1=\seveni
  \textfont2=\twelvesy   \scriptfont2=\tensy   \scriptscriptfont2=\sevensy
  \textfont3=\twelveex   \scriptfont3=\twelveex  \scriptscriptfont3=\twelveex
  \textfont\itfam=\twelveit
  \textfont\slfam=\twelvesl
  \textfont\bffam=\twelvebf \scriptfont\bffam=\tenbf
  \scriptscriptfont\bffam=\sevenbf
  \normalbaselines\rm}



\def\beginlinemode{\endmode
  \begingroup\parskip=0pt \obeylines\def\\{\par}\def\endmode{\par\endgroup}}
\def\beginparmode{\endmode
  \begingroup \def\endmode{\par\endgroup}}
\let\endmode=\par
{\obeylines\gdef\
{}}
\def\singlespace{\baselineskip=\normalbaselineskip}

\def\oneandahalfspace{\baselineskip=\normalbaselineskip
  \multiply\baselineskip by 3 \divide\baselineskip by 2}
\def\doublespace{\baselineskip=\normalbaselineskip \multiply\baselineskip by 2}

\newcount\firstpageno
\firstpageno=2
\footline={\ifnum\pageno<\firstpageno{\hfil}\else{\hfil\twelverm\folio\hfil}\fi}
\def\toppageno{\global\footline={\hfil}\global\headline
  ={\ifnum\pageno<\firstpageno{\hfil}\else{\hfil\twelverm\folio\hfil}\fi}}
\let\rawfootnote=\footnote              
\def\footnote#1#2{{\rm\singlespace\parindent=0pt\parskip=0pt
  \rawfootnote{#1}{#2\hfill\vrule height 0pt depth 6pt width 0pt}}}
\def\raggedcenter{\leftskip=4em plus 12em \rightskip=\leftskip
  \parindent=0pt \parfillskip=0pt \spaceskip=.3333em \xspaceskip=.5em
  \pretolerance=9999 \tolerance=9999
  \hyphenpenalty=9999 \exhyphenpenalty=9999 }
\def\dateline{\rightline{\ifcase\month\or
  January\or February\or March\or April\or May\or June\or
  July\or August\or September\or October\or November\or December\fi
  \space\number\year}}
\def\received{\vskip 3pt plus 0.2fill
 \centerline{\sl (Received\space\ifcase\month\or
  January\or February\or March\or April\or May\or June\or
  July\or August\or September\or October\or November\or December\fi
  \qquad, \number\year)}}


\hsize=6.5truein
\vsize=9.9truein  
\voffset=-1.0truein
\parskip=\medskipamount
\def\\{\cr}
\twelvepoint            
\doublespace            
\overfullrule=0pt       

\def\title                      
  {\null\vskip 3pt plus 0.2fill
   \beginlinemode \doublespace \raggedcenter \bf}

\def\author                     
  {\vskip 3pt plus 0.2fill \beginlinemode
   \singlespace \raggedcenter\sc}

\def\affil                      
  {\vskip 3pt plus 0.1fill \beginlinemode
   \oneandahalfspace \raggedcenter \sl}

\def\abstract                   
  {\vskip 3pt plus 0.3fill \beginparmode
   \singlespace ABSTRACT: }

\def\endtopmatter               
  {\endpage                     
   \body}

\def\body                       
  {\beginparmode}               

\def\head#1{                    
  \goodbreak\vskip 0.5truein    
  {\immediate\write16{#1}
   \raggedcenter \uppercase{#1}\par}
   \nobreak\vskip 0.25truein\nobreak}

\def\beginitems{
\par\medskip\bgroup\def\i##1 {\item{##1}}\def\ii##1 {\itemitem{##1}}
\leftskip=36pt\parskip=0pt}
\def\enditems{\par\egroup}

\def\beneathrel#1\under#2{\mathrel{\mathop{#2}\limits_{#1}}}

\def\refto#1{$^{#1}$}           

\def\references                 
  {\head{References}            
   \beginparmode
   \frenchspacing \parindent=0pt \leftskip=1truecm
   \parskip=8pt plus 3pt \everypar{\hangindent=\parindent}}

\gdef\refis#1{\item{#1.\ }}                     

\gdef\journal#1, #2, #3, 1#4#5#6{               
    {\sl #1~}{\bf #2}, #3 (1#4#5#6)}            

\gdef\refa#1, #2, #3, #4, 1#5#6#7.{\noindent#1, #2 {\bf #3}, #4 (1#5#6#7).\rm}

\gdef\refb#1, #2, #3, #4, 1#5#6#7.{\noindent#1 (1#5#6#7), #2 {\bf #3}, #4.\rm}

\def\endreferences{\body}

\def\endpage                    
  {\vfill\eject}

\def\endpaper                   
  {\endmode\vfill\supereject}

\def\ref#1{Ref.~#1}                     
\def\Ref#1{Ref.~#1}                     
\def\[#1]{[\cite{#1}]}
\def\cite#1{{#1}}
\def\(#1){(\call{#1})}
\def\call#1{{#1}}
\def\taghead#1{}
\def\frac#1#2{{#1 \over #2}}
\def\half{{\frac 12}}

\def\12{{1\over2}}

\catcode`@=11
\newcount\r@fcount \r@fcount=0
\newcount\r@fcurr
\immediate\newwrite\reffile
\newif\ifr@ffile\r@ffilefalse
\def\w@rnwrite#1{\ifr@ffile\immediate\write\reffile{#1}\fi\message{#1}}

\def\writer@f#1>>{}
\def\referencefile{
  \r@ffiletrue\immediate\openout\reffile=\jobname.ref%
  \def\writer@f##1>>{\ifr@ffile\immediate\write\reffile%
    {\noexpand\refis{##1} = \csname r@fnum##1\endcsname = %
     \expandafter\expandafter\expandafter\strip@t\expandafter%
     \meaning\csname r@ftext\csname r@fnum##1\endcsname\endcsname}\fi}%
  \def\strip@t##1>>{}}

\def\citeall#1{\xdef#1##1{#1{\noexpand\cite{##1}}}}
\def\cite#1{\each@rg\citer@nge{#1}}	

\def\each@rg#1#2{{\let\thecsname=#1\expandafter\first@rg#2,\end,}}
\def\first@rg#1,{\thecsname{#1}\apply@rg}	
\def\apply@rg#1,{\ifx\end#1\let\next=\relax
\else,\thecsname{#1}\let\next=\apply@rg\fi\next}

\def\citer@nge#1{\citedor@nge#1-\end-}	
\def\citer@ngeat#1\end-{#1}
\def\citedor@nge#1-#2-{\ifx\end#2\r@featspace#1 
  \else\citel@@p{#1}{#2}\citer@ngeat\fi}	
\def\citel@@p#1#2{\ifnum#1>#2{\errmessage{Reference range #1-#2\space is bad.}%
    \errhelp{If you cite a series of references by the notation M-N, then M and
    N must be integers, and N must be greater than or equal to M.}}\else%
 {\count0=#1\count1=#2\advance\count1 by1\relax\expandafter\r@fcite\the\count0,
  \loop\advance\count0 by1\relax
    \ifnum\count0<\count1,\expandafter\r@fcite\the\count0,%
  \repeat}\fi}

\def\r@featspace#1#2 {\r@fcite#1#2,}	
\def\r@fcite#1,{\ifuncit@d{#1}
    \newr@f{#1}%
    \expandafter\gdef\csname r@ftext\number\r@fcount\endcsname%
                     {\message{Reference #1 to be supplied.}%
                      \writer@f#1>>#1 to be supplied.\par}%
 \fi%
 \csname r@fnum#1\endcsname}
\def\ifuncit@d#1{\expandafter\ifx\csname r@fnum#1\endcsname\relax}%
\def\newr@f#1{\global\advance\r@fcount by1%
    \expandafter\xdef\csname r@fnum#1\endcsname{\number\r@fcount}}

\let\r@fis=\refis			
\def\refis#1#2#3\par{\ifuncit@d{#1}
   \newr@f{#1}%
   \w@rnwrite{Reference #1=\number\r@fcount\space is not cited up to now.}\fi%
  \expandafter\gdef\csname r@ftext\csname r@fnum#1\endcsname\endcsname%
  {\writer@f#1>>#2#3\par}}

\def\ignoreuncited{
   \def\refis##1##2##3\par{\ifuncit@d{##1}%
    \else\expandafter\gdef\csname r@ftext\csname r@fnum##1\endcsname\endcsname%
     {\writer@f##1>>##2##3\par}\fi}}

\def\r@ferr{\endreferences\errmessage{I was expecting to see
\noexpand\endreferences before now;  I have inserted it here.}}
\let\r@ferences=\references
\def\references{\r@ferences\def\endmode{\r@ferr\par\endgroup}}

\let\endr@ferences=\endreferences
\def\endreferences{\r@fcurr=0
  {\loop\ifnum\r@fcurr<\r@fcount
    \advance\r@fcurr by 1\relax\expandafter\r@fis\expandafter{\number\r@fcurr}%
    \csname r@ftext\number\r@fcurr\endcsname%
  \repeat}\gdef\r@ferr{}\endr@ferences}


\let\r@fend=\endpaper\gdef\endpaper{\ifr@ffile
\immediate\write16{Cross References written on []\jobname.REF.}\fi\r@fend}

\catcode`@=12

\citeall\refto		
\citeall\ref		%
\citeall\Ref		%


\def\pp{{\prime\prime}}
\def\x{{\bf x}}
\def\la{{\langle}}
\def\ra{\rangle}
\def\ih{{i \over \hbar}}
\def\S{{\Sigma}}
\def\s{{\sigma}}
\def\C{{\bar C}}

\centerline{\bf An Operator Derivation of the Path Decomposition Expansion}

\vskip 0.5in
\author J. J. Halliwell
\vskip 0.2in
\affil
Theory Group
Blackett Laboratory
Imperial College
London SW7 2BZ
UK
\vskip 0.5in
\centerline {\rm Preprint IC 94--95/41}
\vskip 0.2in
\centerline{\rm June 1995}
\vskip 0.2in
\abstract
{The path decomposition expansion is a path integral technique for
decomposing sums over paths in configuration space into sums over
paths in different spatial regions.  It leads to a decomposition of
the configuration space propagator across arbitrary surfaces in
configuration space. It may be used, for example, in calculations of the
distribution of first crossing times. The original proof relied
heavily on the position representation and in particular on the
properties of path integrals.  In this paper, an elementary proof of
the path decomposition expansion is given using projection
operators. This leads to a version of the path decomposition
expansion more general than the configuration space form
previously given. The path decomposition expansion in momentum space
is given as an example.
}
\endtopmatter

\head{\bf 1. Introduction}

The propagator in non-relativistic quantum mechanics is commonly
represented by a sum over paths:
$$
\eqalignno{
g( \x^{\pp}, t^{\pp} | \x', t' )
&\equiv
\la \x^{\pp} | e^{- \ih H (t^{\pp} -t') } | \x' \ra
&(1.1) \cr
&= \int {\cal D}{ \x(t) } \exp \left( \ih S [ \x (t) ] \right)
&(1.2)\cr}
$$
Here, as usual, $S[\x(t)]$ is the action and the sum is over paths
$ \x (t) $ in a $d$--dimensional configuration space
satisfying the boundary conditions $\x (t') = \x' $,
$ \x (t^{\pp} ) = \x^{\pp} $.
The paths in Eq.(1.2) move {\it forwards} in time. An important consequence
of this feature is the composition property of the propagator: Consider
an intermediate surface labeled by $t$, so $t' < t < t^{\pp}$. Then
because the paths move forwards in time, they intersect the surface labeled
by $t$ once and only once, at a point $\x$, say. The paths summed over may
therefore be partitioned according to the value $\x$ at which they cross
the surface labeled by $t$, and one readily derives the composition
law [\cite{Har2,HO}],
$$
g( \x^{\pp}, t^{\pp} | \x', t' )
= \int d^d \x \ g( \x^{\pp}, t^{\pp} | \x, t )
\ g( \x, t| \x', t' )
\eqno(1.3)
$$

A more complicated story arises in the case of decomposition of the propagator
across general surfaces in configuration space.
The paths $\x (t) $ in configuration space go backwards and forwards
in each of the coordinates $x_1, x_2, \cdots $, and will
generally cross a given surface $ \S $, such as $x_1 = constant$,
many times. The point of
crossing is therefore not well-defined.
However, what is well-defined is the time and location of {\it first} crossing
of a surface $\S$. That is, the paths connecting $\x'$ at time $t'$ to
$\x^{\pp}$ at time $t^{\pp}$ (where $\x'$ and $\x^{\pp}$ lie on
opposite sides of
a surface $\S$) may be partitioned according to the time $t_\s $ and location
$\x_{\s}$ of first crossing.
Corresponding to this partition of the paths is a
decomposition of the propagator
called the path decomposition expansion, or PDX [\cite{HO,AK,vB,S}],
$$
g( \x^{\pp}, t^{\pp} | \x', t' )
= \int_{t'}^{t^{\pp}} dt_{\s} \int_{\S} d^{d-1} \x
\ g( \x^{\pp}, t^{\pp} | \x_{\s}, t_{\s} )
\ {i \hbar \over 2 m} {\bf n} \cdot {\bf \nabla}
g^{(r)} ( \x_{\s}, t_{\s} | \x', t' )
\eqno(1.4)
$$
Here, $ g^{(r)} $ is the
restricted propagator on the side of $\S$ containing $\x'$,
and $ {\bf n}$ is the normal to the surface $\S$ pointing
away from the region of restricted propagation.
The restricted propagator $ g^{(r)} $ is defined to vanish
on $\S$ but its normal derivative does not. Eq.(1.4) consists
of two parts. The first term in the integrand describes unrestricted
propagation from the surface to the final point. The second term
therefore describes the sum over paths which never cross $\S$ but end
on it at $\x_{\s}$ at time $t_{\s}$.
One may also consider the case in which the initial and final points
lie on the same side of $\S$, which leads to the expression,
$$
\eqalignno{
g( \x^{\pp}, t^{\pp} | \x', t' )
= & g^{(r)}( \x^{\pp}, t^{\pp} | \x', t' )
\cr
& + \int_{t'}^{t^{\pp}} dt_{\s} \int_{\S} d^{d-1} \x
\ g( \x^{\pp}, t^{\pp} | \x_{\s}, t_{\s} )
\ {i \hbar \over 2 m} {\bf n} \cdot {\bf \nabla}
g^{(r)} ( \x_{\s}, t_{\s} | \x', t' )
&(1.5) \cr}
$$
where again the normal ${\bf n}$ points away from the region of
restricted propagation.

The PDX was originally introduced in connection with calculations
concerning tunneling [\cite{AK}]. It has since been used to derive the
composition laws of relativistic quantum mechanics from their
path integral representation [\cite{HO}]. It is clearly also be of use
for computing spacetime coarse grainings in non-relativistic
quantum mechanics, {\it i.e.}, probabilities of
certain types of alternatives which cannot be expressed in terms of
a wave function at a single moment of time [\cite{Har,Yam}].

Aan example of a spacetime coarse graining in which the above
formulae are useful is the first crossing time distribution.
The amplitude to start in a state
$\Psi (\x', t')$ with
support on one side of $\S$ only, to cross the surface for the first time
$\S$ in the time interval $ [ t_1, t_2 ] $ (where $t' < t_1 < t_2 < t^{\pp}$),
and then to end up at $\x^{\pp}$ at time $t^{\pp}$. This is given by
$$
\eqalignno{
A (t_1, t_2, \x^{\pp}, t^{\pp} ) =
& \int d^d \x' \int_{t_1}^{t_2} dt_{\s} \int_{\S} d^{d-1} \x
\ g( \x^{\pp}, t^{\pp} | \x_{\s}, t_{\s} )
\cr
& \times \ {i \hbar \over 2 m} {\bf n} \cdot {\bf \nabla}
g^{(r)} ( \x_{\s}, t_{\s} | \x', t' )
\ \Psi (\x', t')
&(1.6) \cr}
$$
The candidate probability of crossing the surface for the first time
in the time interval $[t_1, t_2]$ is therefore
$$
p(t_1, t_2) = \int d^d  \x^{\pp}
\Bigl | A (t_1, t_2, \x^{\pp}, t^{\pp} ) \Bigr |^2
\eqno(1.7)
$$
It is referred to as a candidate probability because the so-called
``probability sum rules'' are not in general satisfied by objects
constructed in this way, and thus (1.7) is not a true probability.
In this case, the rule to be satisfied is that the candidate
probability (1.7) and the probability of never crossing
the surface in the time interval $[t_1, t_2]$ must sum to 1.
This is generally not true unless the initial state is restricted in
some way, or the system is coupled to a wider environment
[\cite{Har,Yam}]. We will not go into this issue here, although it
is often important to keep it in mind.

The original proof of the PDX involved a detailed treatment of the
Euclidean path integral, and relied on a particular
integral identity [\cite{AK}]. A more sophisticated proof was
given in Ref.[\cite{HO}], using a rigorous definition of the
Euclidean sum over histories.
A proof using the configuration space propagator in the
energy representation has also been given [\cite{vB}].
All of these proofs use the configuration space propagator, and
the first two in particular, rely on the notion of sums over paths
in configuration space. However, first crossing questions involving
only position are clearly not the most general. It is
reasonable to ask, for example, for the amplitude of a first
crossing in momentum space.

In this paper, it is shown that the path decomposition expansion may be
proved in a way that minimizes reliance on the properties
of paths in configuration space. The proof uses projection
operators rather than sums over histories.
A form of the PDX is thus obtained
which is valid for first crossings a wide class of observables,
not just position. As an example, the PDX
for the case of first crossing in momentum space is derived.

\head{\bf 2. A New Proof of the Path Decomposition Expansion}

Suppose configuration space is divided into two regions, $C$, and its
complement
$ \C$, and let $\S$ be their common boundary.
We are interested in
the propagator from a point $\x'$ in $\C$ at $t'$ to
$ \x^{\pp}$ in $C$ at $t^{\pp}$. Introduce the projection operator onto
$ C $,
$$
P_C = \int_C d^d \x \ | \x \ra \la \x |
\eqno(2.1)
$$
Its complement $P_\C $ is analogously defined, and we have
the important relations,
$$
\eqalignno{
P_C + P_{\C} &= 1
&(2.2)
\cr
P_C P_{\C} & = 0
&(2.3) \cr}
$$
Introduce the discrete set of times
$ t' = t_0 < t_1 < t_2 < \cdots < t_n  =t^{\pp} $.
We will eventually take the continuum limit, in which
$ (t_k - t_{k-1} ) \rightarrow 0 $, and $n \rightarrow \infty$,
whilst $ t_n - t_0 $ remains constant.
Introduce the Heisenberg picture projections
$$
P(t) = e^{\ih H (t - t_0) } P e^{ - \ih H (t -t_0) }
\eqno(2.4)
$$
so $P(t_0) = P$.

The PDX follows directly from a resolution of the
identity operator which we now derive. Consider the resolution
of the identity (2.2) at time $t_0$.
Multiplying the last term by the same resolution of the identity at
time $t_1$, one obtains
$$
1 = P_C (t_0 ) + P_C (t_1 ) P_{\C} (t_0 ) + P_{\C} (t_1) P_{\C} (t_0)
\eqno(2.5)
$$
Multiplying the last term by the same resolution of the identity at
$ t_2$, and proceeding iteratively leads to the result
$$
\eqalignno{
1 = P_C (t_0) & + \sum_{k=1}^n P_C (t_k) P_{\C} (t_{k-1} ) \cdots P_{\C} (t_0 )
\cr
& + P_{\C} (t_n ) \cdots P_{\C} (t_0 )
&(2.6) \cr }
$$
Each term in this sum corresponds to the statement that the particle
is in $\C$ at times $t_0, t_1 \cdots t_{k-1}$, in $C$ at $t_k$, and
in $\C$ or $C$ at times $t_{k+1}$ to $t_n$. In the continuum limit
each term will therefore represent the statement that the particle
crosses $\S$ for the first time at $t_k$. We cannot of course say
this without taking the continuum limit, because the particle could
be anywhere between each time at which the projection acts.

Now insert the resolution of the identity (2.6) into the expression for the
propagator (1.1). Note first that we have
$$
P_C (t_0) | \x' \ra = 0
\eqno(2.7)
$$
since $ \x' $ is not in $ C $, and
$$
\la \x^{\pp}| e^{- \ih H (t_n - t_0 ) } P_{\C} (t_n) = 0
\eqno(2.8)
$$
since $\x^{\pp}$ is not in $\C $. It follows that
$$
\la \x^{\pp} | e^{- \ih H (t_n -t_0) } | \x' \ra
= \sum_{k=1}^n \ \la \x^{\pp} | e^{- \ih H (t_n -t_0) }
P_C (t_k) P_{\C} (t_{k-1} ) \cdots P_{\C} (t_0) | \x' \ra
\eqno(2.9)
$$
since (2.7) and (2.8) imply that the first and last terms on the
right-hand side of (2.6) do not contribute.
When the time interval $t_k - t_{k-1} = \delta t $ is small,
we have
$$
P_C (t_k) \approx P_C (t_{k-1} ) + \delta t \dot P_C (t_{k-1} ) +
O (\delta t^2)
\eqno(2.10)
$$
Using (2.3), it follows that
$$
\eqalignno{
\la \x^{\pp} | e^{- \ih H (t_n -t_0) } | \x' \ra
= \sum_{k=1}^n \ \delta t \ \la \x^{\pp} | & e^{- \ih H (t_n -t_{k-1}) }
\dot P_C  P_{\C}
e^{ - \ih H (t_{k-1} - t_{k-2} ) }
\cr
\times
& P_{\C} (t_{k-2} ) \cdots P_{\C} (t_0) | \x' \ra
&(2.11) \cr}
$$
This is now conveniently written,
$$
\la \x^{\pp} | e^{- \ih H (t_n -t_0) } | x' \ra
= \sum_{k=1}^n \ \delta t \ \la \phi | \dot P_C | \chi \ra
\eqno(2.12)
$$
where
$$
\eqalignno{
\la \phi | \x \ra &= \phi^* (\x) =
\la \x^{\pp} | e^{- \ih H (t_n -t_{k-1}) } | \x \ra
&(2.13)
\cr
\la \x | \chi \ra &= \chi (\x) = \la \x | P_{\C}
e^{ - \ih H (t_{k-1} - t_{k-2} ) }
P_{\C} (t_{k-2} ) \cdots P_{\C} (t_0) | \x' \ra
&(2.14)
\cr }
$$
Eq.(2.13) is clearly the propagator
from $(\x, t_{k-1})$ to $ (\x^{\pp}, t_n) $.
In the continuum limit $\delta t \rightarrow 0$,
Eq.(2.14) becomes the restricted propagator from $\x '$ to $\x$
in the region $ \C $, and vanishes on the boundary $\S$
between $C$ and $\C$.

This expression is readily simplified. Suppose the
Hamiltonian is
$$
H = { {\bf p}^2 \over 2m} + V(\x)
\eqno(2.15)
$$
Then
$$
\dot P_C = \ih\ [ H, P_C ] = \ih \ [{ {\bf p}^2 \over 2m}, P_C ]
\eqno(2.16)
$$
and it follows that
$$
\eqalignno{
\la \phi | \dot P_C | \chi \ra
&= \ih \int_C d^d x \ \left( - { \hbar^2 \over 2m}
\chi \nabla^2 \phi^*  + { \hbar^2 \over 2m} \phi^*
 \nabla^2 \chi  \right)
\cr &=
- { i \hbar \over 2m} \int_\S d^{d-1} x \ {\bf n}
\cdot \left( \chi {\bf \nabla} \phi^* - \phi^*
{\bf \nabla} \chi \right)
&(2.17) \cr }
$$
Now taking the continuum limit, the discrete sum becomes
an integral, and $ \la \x | \chi \ra $ vanishes on $\S$.
Denoting $t_k$ by $t_{\s}$, we thus obtain
$$
\la \x^{\pp} | e^{- \ih H (t^{\pp} -t') } | \x' \ra
= \int_{t'}^{t^{\pp}} dt_{\s} \int_{\S} d^{d-1} \x
\ \phi^* (\x) {i \hbar \over 2 m} {\bf n} \cdot {\bf \nabla}
\chi (\x)
\eqno(2.18)
$$
Inserting (2.13) and (2.14) yields the desired result, Eq.(1.4).
If $\x '$ and $\x^{\pp}$ are on the same side of $\S$, in the region
$\C$ say, then the last term on the right-hand side of Eq.(2.6)
also contributes, and the result (1.5) is obtained.

We have therefore derived the PDX in a way that appealed to the
properties of the position operator only in the final steps, (2.16), (2.17).
This points the way to a form of the PDX which should be valid for
any projections onto any observable (provided the appropriate restricted
propagators exist).
In particular, the continuum limit of Eq.(2.6), multiplied by the
unitary evolution operator yields
$$
\eqalignno{
e^{- \ih H (t^{\pp} -t') }
=& \ e^{- \ih H (t^{\pp} -t') } \ P_C
\cr
& + \int_{t'}^{t^{\pp}} dt_{\s} \ e^{- \ih H (t^{\pp} -t_{\s} ) }
\ \ih [ H, P_C ] \ G^{(r)}(t_{\s},t')
\cr
& + G^{(r)} (t^{\pp}, t')
&(2.19) \cr}
$$
where
$$
G^{(r)}(t_{\s},t') = {\rm lim} \ e^{- \ih H (t_{\s} -t') }
\ P_{\C} (t_{k-1} )
P_{\C} (t_{k-2} ) \cdots P_{\C} (t_0)
\eqno(2.20)
$$
and similarly for $ G^{(r)} (t^{\pp}, t') $, where
the limit is $ \delta t \rightarrow 0 $, $ k \rightarrow \infty $
with $t_k -t_0 $ held constant. Clearly the first term in (2.19)
does not contribute for initial states with non-zero support in
$\C$ only. Eq.(2.19), a generalization of the PDX,
is the main result of this paper.

\head{\bf 3. First Crossing in Momentum Space}

As an example of the generalized PDX,
consider the case of first crossing in momentum space.
For simplicity let the system be one-dimensional, and let the
region $C$ be $p>0$, so $\S$ is the surface $p=0$.
Then
$$
P_C = \int_0^{\infty} dp \ | p \ra \la p |
\eqno(3.1)
$$
The PDX has the form
$$
\la p^{\pp} | e^{ - \ih H (t^{\pp} -t') } | p' \ra
=
\int_{t'}^{t^{\pp}} dt_{\s}
\ \la \phi | \dot P_C | \chi \ra
\eqno(3.2)
$$
where
$$
\la \phi | p \ra =
\la p^{\pp} | e^{ - \ih H (t^{\pp} -t_{\s} ) } | p \ra
= g( p^{\pp}, t^{\pp} | p, t' )
\eqno(3.3)
$$
and
$$
\la p | \chi \ra = g^{(r)} ( p, t_{\s} | p', t' )
\eqno(3.4)
$$
Eq.(3.4) is the propagator in momentum space restricted to the
region $ p < 0 $ and vanishes on $ p = 0$.

We have
$$
\la \phi | \dot P_C | \chi \ra
= \ih \la \phi | [ V(x), P_C ] | \chi \ra
\eqno(3.5)
$$
For simplicity, let $V(x) = \half m \omega^2 x^2 $. Then following
steps closely analagous to those used in Eq.(2.17), it is readily
seen that
$$
\la \phi | \dot P_C | \chi \ra
= - { i m \omega^2 \hbar \over 2} \left( \phi^* (p)
{ \partial \chi (p)  \over \partial p}
- { \partial \phi^* (p) \over \partial p} \chi (p)
\right)_{p=0}
\eqno(3.6)
$$
Since $ \chi (p) $ vanishes at $p=0$, we derive the PDX in
momentum space,
$$
g(p^{\pp}, t^{\pp} | p', t' )
= - { i m \omega^2 \hbar \over 2}
\int_{t'}^{t^{\pp}} dt_{\s}
\ g(p^{\pp}, t^{\pp} | p=0, t_{\s})
\ { \partial g^{(r)} \over \partial p} (p=0, t_{\s} | p', t')
\eqno(3.7)
$$

Other examples of the generalized PDX are easily constructed.
Eq.(2.19) is valid for spin systems, for example, although it is
less clear how useful it might be there. These and similar
considerations will be pursued elsewhere.

\head{\bf Acknowledgements}

I am grateful to Don Marolf for useful conversations and a critical
reading of the manuscript.

\references

\def\np{{\sl Nucl. Phys.\ }}

\refis{AK} A. Auerbach and S. Kivelson, \np {\bf B257}, 799 (1985).

\refis{vB} P. van Baal, ``Tunneling and the path decomposition expansion'',
Utrecht Preprint THU-91/19 (1991).

\refis{HO} J. J. Halliwell and M. E. Ortiz, {\sl Phys.Rev.} {\bf D48}, 748
(1993).

\refis{Har} J. B. Hartle, {\sl Phys.Rev.} {\bf D44}, 3173 (1991).

\refis{Har2} J. B. Hartle, {\sl Phys.Rev.} {\bf D37}, 2818 (1988).

\refis{S} L. Schulman and R. W. Ziolkowiski, in {\sl Path integrals from
meV to MeV}, edited by V. Sa-yakanit, W. Sritrakool, J. Berananda, M. C.
Gutzwiller, A. Inomata, S. Lundqvist, J. R. Klauder and L. S. Schulman
(World Scientific, Singapore, 1989).

\refis{Yam} N. Yamada and S. Takagi, {\sl Prog.Theor.Phys.}
{\bf 85}, 985 (1991); {\bf 86}, 599 (1991); {\bf 87}, 77 (1992);
N. Yamada, {\sl Sci. Rep. T\^ohoku Uni., Series 8}, {\bf 12}, 177
(1992).

\endreferences

\end